\documentclass{emulateapj}
\usepackage{apjfonts}

\slugcomment{Submitted to ApJ Letters on 3/19/2008; accepted on 11/08/2008}

\shortauthors{Spoon et al.}

\begin{document}

\title{High-velocity neon line emission from the ULIRG IRAS\,F00183--7111: \\ 
revealing the optically obscured base of a nuclear outflow}

\author{H.W.W. Spoon\altaffilmark{1}}
\email{spoon@isc.astro.cornell.edu}
\author{L. Armus\altaffilmark{2}}
\author{J.A. Marshall\altaffilmark{3,4}}
\author{J. Bernard-Salas\altaffilmark{1}}
\author{D. Farrah\altaffilmark{1}}
\author{V. Charmandaris\altaffilmark{5,6}}
\author{B.R. Kent\altaffilmark{7}}

\altaffiltext{1}{Cornell University, Astronomy Department, Ithaca, NY 14853}
\altaffiltext{2}{California Institute of Technology, Spitzer Science Center, 
                 MS 220-6, Pasadena, CA 91125}
\altaffiltext{3}{California Institute of Technology, Pasadena, CA 91125}
\altaffiltext{4}{Jet Propulsion Laboratory, 4800 Oak Grove Drive, 
                 Pasadena, CA 91109}
\altaffiltext{5}{Department of Physics, University of Crete,   
                 GR-71003, Heraklion, Greece}
\altaffiltext{6}{IESL / Foundation for Research and Technology-Hellas, 
                 PO Box 1527, GR-71110, Heraklion, Greece, and Chercheur 
                 Associ\'e, Observatoire de Paris, F-75014,  Paris, France}
\altaffiltext{7}{Jansky Fellow of the NRAO, 520 Edgemont Road, Charlottesville, VA 22901}

\begin{abstract}
We report the first mid-infrared detection of highly disturbed 
ionized gas in the ultraluminous infrared galaxy (ULIRG) 
IRAS\,F00183--7111. The gas, traced by the 12.81\,$\mu$m 
[Ne {\sc ii}] and 15.56\,$\mu$m [Ne {\sc iii}] lines, 
spans a velocity range of
--3500 to +3000 km s$^{-1}$ with respect to systemic velocity.
Optical and near-infrared spectroscopic studies show no 
evidence for similarly high velocity gas components in 
forbidden lines at shorter wavelengths. We interpret this 
as the result of strong extinction (A$_V$=10--50) on the
high-velocity gas, which identifies the base of the outflow 
traced in 5007\,\AA\ [O {\sc iii}] as a plausible origin.
Unusual excitation conditions are implied by a comparison
of the mid-infrared low-excitation neon line emission and
the Polycyclic Aromatic Hydrocarbon (PAH) emission for a
sample of 56 ULIRGs. For IRAS\,F00183--7111, the neon/PAH 
ratio is 8 times higher than the average ratio.
Similar mid-infrared kinematic and excitation characteristics 
are found for only two other ULIRGs in our sample: 
IRAS\,12127--1412NE and IRAS\,13451+1232. Both sources have 
an elevated neon/PAH ratio and exhibit pronounced 
blue wings in their 15.56\,$\mu$m [Ne {\sc iii}] line 
profiles. IRAS\,13451+1232 even shows a strongly blue 
shifted and broad 14.32\,$\mu$m [Ne {\sc v}] line.
While for IRAS\,13451+1232 the observed 
[Ne {\sc iii}]/[Ne {\sc ii}] and [Ne {\sc v}]/[Ne {\sc ii}] 
line ratios indicate exposure of the blue shifted gas 
to direct radiation from the AGN, for IRAS\,F00183--7111 
and IRAS\,12127--1412NE the neon line ratios are consistent
with an origin in fast shocks in a high-density environment,
and with an evolutionary scenario in which strongly blue 
shifted [Ne {\sc ii}] and [Ne {\sc iii}] emission  
trace the (partial) disruption of the obscuring medium 
around a buried AGN. The detection of strongly blue 
shifted [Ne {\sc v}] emission in IRAS\,13451+1232 would 
then indicate this process to be much further advanced 
in this ULIRG than in IRAS\,F00183--7111 and 
IRAS\,12127--1412NE, where this line is undetected.
\end{abstract}


\section{Introduction}

Large-scale galactic outflows of gas (sometimes called 
"superwinds") are seen in galaxies both at low and high 
redshifts \citep[e.g.][]{veilleux05,pettini02}. 
Superwinds are generated when the kinetic energy in 
the outflows from massive stars and supernovae is thermalized, 
generating a region of very hot (T$\sim$10$^6$--10$^7$\,K) 
low-density gas in the ISM of a starburst galaxy 
\citep{chevalier85}. 
As the bubble breaks out of the disk of the galaxy, it 
will rupture, producing a weakly-collimated bipolar outflow 
into the galaxy halo. As the bubble expands it sweeps up and 
shocks ambient material, creating an outflowing wind 
\citep{castor75,weaver77}.
Superwinds may also be driven, at least in part, by radiation 
pressure on dust grains \citep{murray05}.
They may also be driven directly by the interaction of jets
on the dense ISM surrounding an AGN \citep[e.g.][]{holt03}.

There is morphological, physical and kinematic evidence for 
superwinds in nearby starburst and infrared luminous galaxies 
\citep[e.g.][]{heckman90}. Large-scale optical emission-line 
and associated X-ray nebulae are ubiquitous in starbursts
\citep{armus90,lehnert99,strickland04}, 
and these increase in size and luminosity from dwarf starbursts 
all the way to ULIRGs (where they can be tens of Kpc in size ---
\citep{grimes05}. In addition, the winds are 
often seen via blue-shifted absorption features (e.g. Na\,D) 
against the starburst nuclei \citep{heckman00,rupke05,martin05}
indicating that the neutral, cold ISM is being swept-up in 
the outflow. At high redshifts, superwinds have been seen in 
Lyman break galaxies at z=2--3 \citep{pettini01,shapley03}.

The ultraluminous infrared galaxy (ULIRG: 
L(8--1000\,$\mu$m)=10$^{12-13}$ L$_{\odot}$) 
IRAS\,F00183--7111 is known from optical emission
line spectroscopy to harbour such a galactic-scale outflow 
\citep{heckman90}. In this paper, we study the 
mid-infrared evidence for this outflow based on the 
{\it Spitzer}-IRS emission line spectroscopy. The emission 
line data is analyzed and presented along with 
new 3.6, 5.8, 70 and 160\,$\mu$m {\it Spitzer} photometric 
observations, obtained to better constrain the infrared 
spectral energy distribution of this interesting ULIRG.

\begin{figure}
\begin{center}
\includegraphics[angle=0,width=85mm]{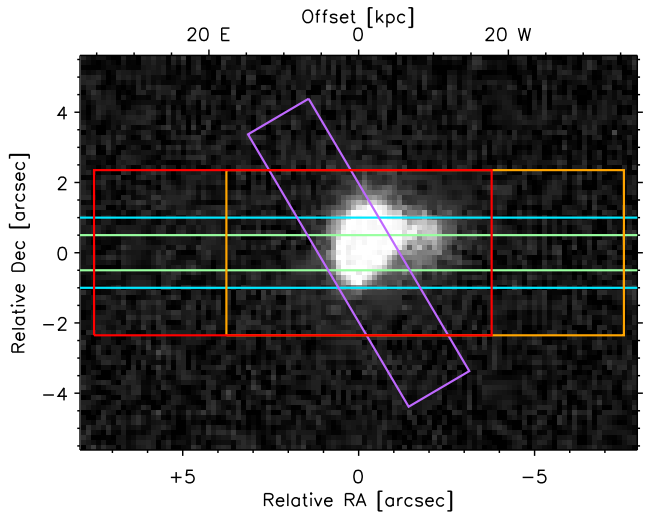}
\caption{K-band image of IRAS\,F00183--7111 replotted from 
\citet{rigopoulou99}. North is up, East is to the left. Overlaid 
on the image are the various slits used to obtain spectra of 
the emission line features discussed in the paper. In {\it red}
and {\it orange} is the {\it Spitzer}-IRS SH slit in the two
nodding positions; in {\it blue} is the slit used by 
\citet{heckman90}, in {\it green} the slit used by
\citet{dannerbauer05}, and in {\it purple} the slit
used by \citet{buchanan06}
\label{fig7}}
\end{center}
\end{figure}

With an integrated 8--1000\,$\mu$m luminosity of 
8.6$\pm$1.3$\times$10$^{12}$ L$_{\odot}$ (this work) 
IRAS\,F00183--7111 is one of the most luminous ULIRGs 
discovered by IRAS.
Near-infrared imaging by \citet{rigopoulou99} (Fig.\,\ref{fig7}) 
shows a disturbed morphology and what appears to be a single nucleus.
Evidence for the presence of an accreting central supermassive 
black hole is provided by the radio luminosity, which is in the 
range of powerful radio galaxies and a factor ten in excess
of what would be expected for starburst galaxies based on the 
far-infrared luminosity \citep[q=1.14;][]{roy97,norris88}. 
The presence of an AGN was recently confirmed by \citet{nandra07}, 
who detected a 6.7\,keV Fe\,K$\alpha$ line (Fe\,{\sc xxv}) 
with a large equivalent width, indicative of reflected light 
from a compton thick AGN.

The most common 
star formation tracers, the PAH emission bands, were not detected 
in the 5--12\,$\mu$m ISO-CAM-CVF spectrum \citep{tran01}. Instead, 
the spectrum is dominated by a strong silicate absorption band 
centered at 9.7\,$\mu$m, indicative of strong absorption of a
background continuum source, most likely a deeply buried AGN 
\citep{tran01}. One of the PAH emission bands, the C--H bending 
mode at 11.2\,$\mu$m, was later identified in the Spitzer-IRS
low-resolution spectrum and was used to constrain the contribution
of mid-infrared detectable star formation to the energy budget of 
IRAS\,F00183--7111 to be $\leq$30\% \citep{spoon04}. The same 
spectrum may also offer indications for ongoing {\it buried} 
star formation through the detection of absorption bands of
crystalline silicate features at 16, 19 and 23\,$\mu$m 
\citep{spoon06}. Crystalline silicates are known to be forged 
by young stars in their circumstellar disks and by evolved 
stars in their dense stellar winds \cite[][and references therein]{molster05}.
Their presence in the general ISM of IRAS\,F00183--7111 may 
indicate fierce ongoing star formation at mid-infrared 
optical depths significantly larger than one \citep{spoon06}.

Optical long-slit spectroscopy of IRAS\,F00183--7111 has 
revealed the presence of kinematically highly disturbed 
ionized gas extending from 10\,kpc West to $\sim$25\,kpc East 
of the nucleus. This gas, traced in both the 4959 and 5008\AA\
[O {\sc iii}] lines, is clearly blue shifted between
5 and 20\,kpc East and shows line widths in excess of
600\,km s$^{-1}$ (FWHM) over a range of 25\,kpc. 
The double peaked profile of both [O {\sc iii}] lines
in this range is in agreement with a superwind model 
in which the line emission originates from the thin 
outer shell of shock-heated, swept-up ambient gas.
Beyond 25\,kpc East, the gas is much more
quiescent, with line widths of $\sim$200\,km s$^{-1}$ 
centered close to systemic velocity \citep{heckman90}. 

Recently, \citet{nandra07} have pointed out that the presence 
of an outflow may also explain the clear eastward offset by 
about 22\,kpc of the soft (0.3--1.0\,keV) X-ray radiation
detected from IRAS\,F00183--7111. It also may identify the
secondary peak at 25\,kpc East in the R-band image\footnote{at 
the redshift of the source the R-band filter includes the 5008\AA\ 
[O {\sc iii}] line \citep{drake04}} by \cite{drake04} 
to be related to this outflow \citep{nandra07}. 
A corresponding K-band peak is absent in the 2.2\,$\mu$m image 
(rest frame 1.65\,$\mu$m) by \cite{rigopoulou99} (Fig.\,\ref{fig7}).

In this paper we report the first identification in a ULIRG
of broad, asymmetric and somewhat blue shifted emission lines of 
[Ne {\sc ii}] at 12.81\,$\mu$m and [Ne {\sc iii}] at 15.56\,$\mu$m
in the high-resolution {\it Spitzer}-IRS spectrum of 
IRAS\,F00183--7111. We compare our detection to available 
optical and near-infrared spectroscopy to investigate the 
processes responsible for this broadening.
Our paper is organized as follows. In Section\,2 we present 
our observations of IRAS\,F00183--7111. In Section\,3 we present 
the 1--120\,$\mu$m SED and an analysis of the mid-infrared 
emission line spectrum. In Section\,4 we analyze line profiles 
of two other ULIRGs which display strong blue shifts in their
fine-structure neon line profiles. Results are discussed in 
Section\,5 and conclusions are provided in Section\,6. 
Additional results for IRAS F00183--7111 are presented in
the Appendix.
Throughout this paper we assume H$_0$ = 71 km s$^{-1}$ Mpc$^{-1}$, 
$\Omega_M$ = 0.27, $\Omega_\Lambda$ = 0.73 and  $\Omega_K$ = 0.

\begin{deluxetable}{llc}
\tablecolumns{3} 
\tablewidth{0pc}
\tablecaption{Photometry of IRAS\,F00183--7111\label{tab2}}
\tablehead{\colhead{Band} & \colhead{$\lambda_{\rm rest}$} & \colhead{Flux density} \\
           \colhead{}     & \colhead{($\mu$m)}           & \colhead{(mJy)}}
\startdata
IRAC-1   & 2.7               & 2.9 $\pm$ 0.3 \\
IRAC-3   & 4.4               & 22  $\pm$ 2 \\
IRAS-12  & 9.0               & $<$60\tablenotemark{a} \\
IRAS-25  & 18.8              & 133  $\pm$  10\tablenotemark{a} \\
IRAS-60  & 45                & 1200 $\pm$ 837\tablenotemark{a} \\
MIPS-70  & 53                & 1500 $\pm$ 225 \\
IRAS-100 & 75                & 1190 $\pm$ 119\tablenotemark{a} \\
MIPS-160 & 121               &  540 $\pm$ 80 \\
\enddata
\tablenotetext{a}{IRAS Faint Source Catalog v2}
\end{deluxetable}

\section{Observations and data reduction}

\subsection{IRAC and MIPS photometry}

{\it Spitzer} Space Telescope \citep{werner04} IRAC \citep{fazio04} 
3.6 and 5.8\,$\mu$m photometry was obtained on 2006 August 14 
(AOR key 17517824). IRAS\,F00183--7111 was imaged at 
5 dither positions using 2\,sec ramp times. The source flux density 
was measured by integrating over a 12'' radius aperture placed upon 
the SSC pipeline-produced (S14) post-BCD mosaic. We estimate the 
uncertainty in the photometry due to the background to be $\sim$2\%. 
The absolute calibration uncertainties of the IRAC post-BCD images 
are 10\%.
As part of the same program we obtained MIPS \citep{rieke04} 
70 and 160\,$\mu$m photometry on 2006 June 17 (AOR key 17528320). 
IRAS\,F00183--7111 was
imaged for 1 cycle at 70\,$\mu$m and for 4 cycles at 160\,$\mu$m
using 3\,sec ramps. The source flux densities were measured
from the SSC pipeline-produced (S14) unfiltered post-BCD images by
integrating the flux within 8 and 20" radius apertures at 70 and
160\,$\mu$m, respectively. Aperture corrections were applied to
recover the full point-source flux densities. We estimate the 
uncertainty on the 70 and 160\,$\mu$m flux densities due to the 
background to be around 1 and 2\%, respectively. The absolute 
calibration uncertainties of the MIPS post-BCD images are 10--20\%.
We present the results of our photometric observations in 
Table\,\ref{tab2}. The calculation of infrared luminosities
which utilize this photometry are described in Section\,3.1.

\begin{figure}
\begin{center}
\includegraphics[angle=0,width=85mm]{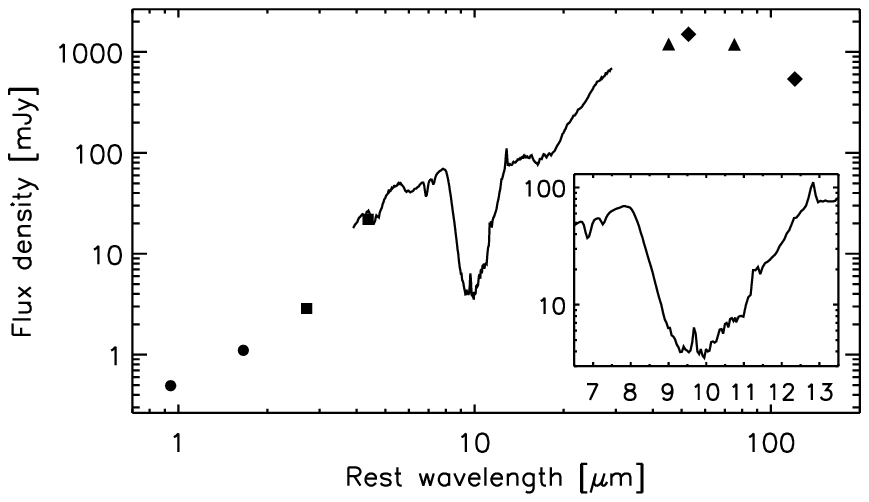}
\caption{The near-to-far-infrared spectral energy distribution
of IRAS\,F00183--7111. The 4--29\,$\mu$m {\it Spitzer}-IRS 
low-resolution spectrum \citep{spoon04} is shown as a 
continuous line. 
Photometric points are labeled with the following symbols:
{\it circles}: ESO-NTT J \& K \citep{rigopoulou99}, 
{\it squares}: IRAC 3.6 \& 5.8\,$\mu$m, {\it triangles}: 
IRAS 60 \& 100\,$\mu$m, {\it diamonds}: MIPS 70 \& 160\,$\mu$m.
Error bars on the photometric points are too small to be
visible. Inset: detailed view of the spectral range around the 
9.7\,$\mu$m silicate absorption profile. Note the emission
lines of H$_2$ S(3) at 9.66\,$\mu$m and [Ne {\sc ii}] at 
12.81\,$\mu$m.
\label{fig4}}
\end{center}
\end{figure}

\subsection{IRS spectroscopy}

High resolution ($\Delta$$v\sim$500 km s$^{-1}$; R$\sim$600) 
10--37\,$\mu$m IRS \citep{houck04} spectra of IRAS\,F00183--7111 
were obtained at two epochs: 2005 June 2 (AOR key 6651904) and 
2008 June 1 (AOR keys 25289472 and 25289984). In both cases the 
source was acquired using a high-accuracy blue peak-up and observed 
in staring mode. The 10--19.5\,$\mu$m portion of the spectrum
was obtained in 9 cycles of 120\,sec integrations in IRS Short High 
(SH; slit size 4.7$"\times$11$"$) at PA=91$\degr$.
The 19.3--37.0\,$\mu$m portion of the spectrum was obtained in
2 sets of 35 cycles of 60\,sec integrations, alternated with 
sky background observations (30 cycles of 60\,sec), in IRS
Long High (LH; slit size 11.1$"\times$22.3$"$) at PA=-15$\degr$.
Figure\,\ref{fig7} shows the SH slit overplotted on the K-band image 
of the source \citep{rigopoulou99}. The LH slit is not shown as it is
more than twice as wide and long as the SH slit.

Data reduction started from 2D {\tt droop} level (i.e. pre-flatfield) 
images provided by the S17.2 pipeline of the {\it Spitzer} Science Center
(SSC). For SH, in a first step we used {\tt IRSCLEAN} to interpolate
over the bad and 'rogue' pixels. For LH, this step was preceeded by 
subtraction of the contemporaneous sky background images. We then used 
the {\tt SMART} data reduction package 
\citep{higdon04} to extract the 1D target spectrum in `full slit' mode.
The same procedure was followed for a large number of observations
of the calibration star $\xi$\,Dra. The final spectrum was obtained 
by multiplying the spectrum of IRAS\,F00183--7111 by the relative
spectral response function created from the ratio of the observed 
spectrum of $\xi$\,Dra to the stellar reference spectrum 
\citep[][G.C. Sloan priv.comm.]{cohen03}. Note that given the
absence of contemporaneous SH background observations, the SH 
portion of the spectrum is {\it not} background-subtracted.

The wavelength calibration of our spectrum is accurate to about
1/5 (SSC, priv. comm.) of a resolution element (500 km s$^{-1}$), 
which amounts to 100 km s$^{-1}$.

\begin{deluxetable}{llcc}
\tablecaption{Synthetic photometry for IRAS\,F00183--7111\label{tab4}}
\tablewidth{0pc} 
\tablehead{\colhead{Band} & \colhead{$\lambda_{\rm rest}$} &
  \colhead{Flux density}  & \colhead{Uncertainty} \\
  \colhead{} & \colhead{($\mu$m)} & \colhead{(mJy)} &  
          }
\startdata
IRAC-1   & 3.6               & 8.6    & 1\% \\
IRAC-2   & 4.5               & 29.6   & 1\% \\
IRAC-3   & 5.8               & 47.3   & 1\% \\
IRAC-4   & 8.0               & 48.6   & 1\% \\ 
MIPS-24  & 24                & 370    & 1\% \\
IRAS-25  & 25                & 360    & 1\% \\
IRAS-60  & 60                & 1410   & 1\% \\
MIPS-70  & 70                & 1190   & 2\% \\
IRAS-100 & 100               & 890    & 3\% \\
MIPS-160 & 160               & 490    & 5\% \\
SCUBA-450& 450               & 20     & 6\% \\
SCUBA-850& 850               & 2.0    & 6\% \\
\enddata
\tablenotetext{}{{\sc Notes}.--- Rest-frame synthetic fluxes measured 
from the model spectrum of IRAS\,F00183--7111 
(see Section\,3.1). The statistical uncertainty in the synthetic
photometry is computed from the uncertainty in the total fitted flux}
\end{deluxetable}


\section{Analysis}

\begin{figure*}
\begin{center}
\plotone{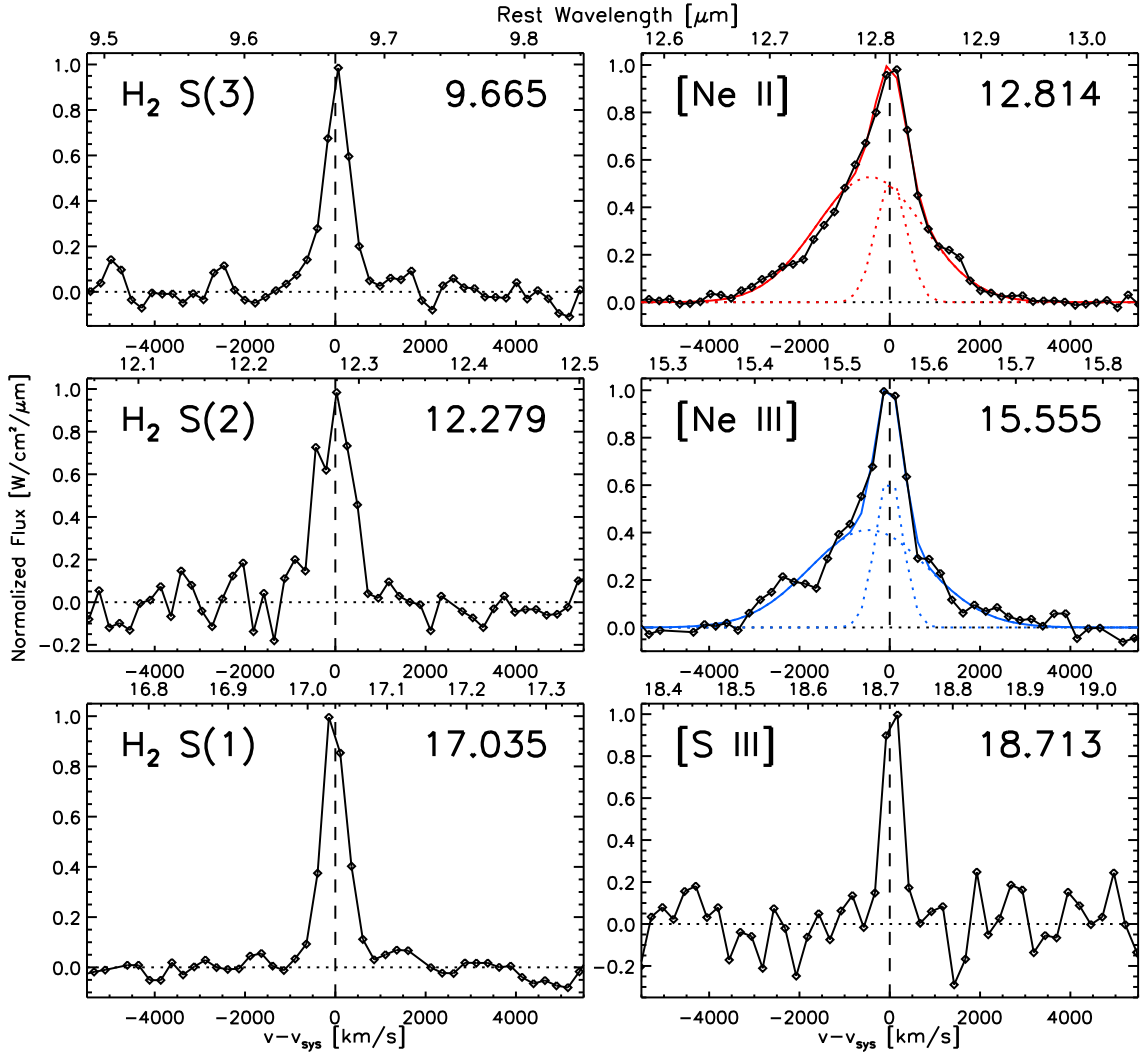}
\caption{Comparison of the emission line profiles for the 
six detected emission lines in the spectrum of IRAS\,F00183--7111.
The left column shows the S(1), S(2) and S(3) pure rotational H$_2$ 
lines, the right panel the $[$Ne {\sc ii}$]$, $[$Ne {\sc iii}$]$ 
and $[$S {\sc iii}$]$ lines. All six lines have been observed 
at the same spectral resolution $R$=600. 
The $[$Ne {\sc ii}$]$ and $[$Ne {\sc iii}$]$ line profiles are 
clearly asymmetric and broadened beyond the instrumental spectral 
profile (FWHM = 500\,km s$^{-1}$). Their profiles are fit by 
two Gaussian components ({\it red} and {\it blue dotted} lines, 
respectively), the sum of which is shown as a {\it red} or 
{\it blue continuous} line
\label{fig1}}
\end{center}
\end{figure*}

\subsection{Infrared SED}

With the addition of the IRAC 3.6\,$\mu$m and 
MIPS 70 and 160\,$\mu$m photometry it is possible
to fit the infrared SED of IRAS\,F00183--7111 over 
the entire rest frame wavelength range of 1--120\,$\mu$m. 
Especially important is the addition of the
MIPS 160\,$\mu$m point, which extends the sampling 
of the rest frame SED from 75\,$\mu$m (pre-{\it Spitzer})
to 120\,$\mu$m (see Fig.\ref{fig4}), allowing 
a more reliable extrapolation of the SED into the 
submillimetre range. For the latter, we have used 
the multi-component spectral energy distribution 
decomposition code of \citet{marshall07}.

Based on our fit, we have computed the following 
integrated rest frame infrared luminosities for 
IRAS\,F00183-7111:
L(1--1000\,$\mu$m)=9.9$\pm$1.3$\times$10$^{12}$ L$_{\odot}$,
L(8--1000\,$\mu$m)=8.6$\pm$1.3$\times$10$^{12}$ L$_{\odot}$,
L(40--500\,$\mu$m)=4.4$\pm$0.8$\times$10$^{12}$ L$_{\odot}$ and
L(40--120\,$\mu$m)=4.0$\pm$0.6$\times$10$^{12}$ L$_{\odot}$.
We have further used the results of the fit to derive 
rest frame filter flux densities for the IRAC, MIPS,
IRAS and SCUBA bands. 
These values are tabulated in Table\,\ref{tab4}. 
Given the 15\% absolute calibration uncertainty 
of the MIPS 70 and 160\,$\mu$m data, and the fact 
that the MIPS 160\,$\mu$m data is the longest 
wavelength point constraining our fit, there is 
clearly some uncertainty in our extrapolation of 
the flux density out to 1000\,$\mu$m. The spectral 
shape of the extrapolated flux density beyond  
160\,$\mu$m is a function of the chosen power-law 
spectral index of the far-IR dust opacity. The 
above stated uncertainties in the integrated 
luminosities incorporate both the systematic 
uncertainty resulting from our choice of this 
dust opacity spectral index as well as the 
statistical uncertainty resulting from the MIPS 
calibration uncertainty.

The ratio of the observed IRAS-25 to IRAS-60 filter 
fluxes is commonly used to classify ULIRGs as "warm"
(25/60\,$\mu$m $>$ 0.2) or "cold" 
(25/60\,$\mu$m $<$ 0.2), separating ULIRGs
believed to be AGN-dominated from those which
are thought to be powered predominantly by star 
formation \citep{degrijp85,sanders88}. For the
classification to be meaningful also for higher
redshift targets like IRAS\,F00183--7111, rest 
frame synthetic IRAS fluxes should be used. For 
IRAS\,F00183--7111, this results in an IRAS-25 
to IRAS-60 flux ratio of 0.26, which places 
it among the class of "warm", AGN-dominated 
ULIRGs, in agreement with results obtained 
at other wavelengths (see Section 1).

\subsection{Emission lines}

The high-resolution {\it Spitzer}-IRS spectrum of IRAS\,F00183--7111 
covers the rest frame wavelength range of 7.5--27.8\,$\mu$m.
In Table\,\ref{tab1} we list the line fluxes and 3$\sigma$ upper
limits for 17 emission lines, as well as for the 11.2\,$\mu$m 
PAH emission feature. For the latter, the feature flux was
measured above a spline continuum defined shortward of 
11.1\,$\mu$m and longward of 11.6\,$\mu$m.
The only lines that are detected in our spectrum are the 
S(1), S(2) and S(3) pure rotational lines of molecular hydrogen,
and the 12.81\,$\mu$m [Ne {\sc ii}], 15.56\,$\mu$m [Ne {\sc iii}] 
and 18.71\,$\mu$m [S {\sc iii}] fine-structure lines. 
In Fig.\,\ref{fig1} we present their continuum-subtracted line 
profiles as a function of velocity relative to systemic\footnote{
In the absence of CO or 21cm redshift determinations, we have 
used the mid-infrared S(1) and S(3) molecular hydrogen lines 
to determine the recession velocity of IRAS\,F00183--7111 
to be 98400$\pm$200\,km s$^{-1}$ (z=0.328 $\pm$ 0.001). This
redshift is 360\,km s$^{-1}$ higher than the value adopted 
by \cite{heckman90} and 180\,km s$^{-1}$ higher than the
value used by \cite{roy97}}.

\begin{deluxetable}{llccc}
\tablecolumns{5} 
\tablewidth{0pc} 
\tablecaption{Emission features in the spectrum of IRAS\,F00183--7111\label{tab1}}
\tablehead{\colhead{Feature ID} & \colhead{$\lambda_{\rm rest}$} & \colhead{Flux} &
           \colhead{obs. FWHM} & \colhead{true FWHM}\\
           \colhead{} & \colhead{($\mu$m)} & \colhead{(10$^{-21}$ W cm$^{-2}$)} & 
           \colhead{(km s$^{-1}$)} & \colhead{(km s$^{-1}$)}
          }
\startdata
$[$Ne \sc{vi}$]$  & 7.65               & $<$0.25                & ... & ...\\
H$_2$ S(4)        & 8.03               & $<$0.26                & ... & ...\\
$[$Ar \sc{iii}$]$ & 8.99               & $<$0.07                & ... & ...\\
H$_2$ S(3)        & 9.66               & 0.52 $\pm$ 0.03        & 650$\pm$40 &410$\pm$50\\ 
$[$S \sc{iv}$]$   & 10.51              & $<$0.08                & ... & ...\\
PAH               & 11.2               & 2.7 $\pm$ 0.4          & ... & ...\\
H$_2$ S(2)        & 12.28              & 0.59 $\pm$ 0.1         & 950$\pm$130 &810$\pm$120\\
Hu$\alpha$        & 12.37              & $<$0.12                & ... & ...\\
$[$Ne \sc{ii}$]$  & 12.81              & 6.3 $\pm$ 0.3          &1500$\pm$100 &1410$\pm$100\\
$[$Ne \sc{v}$]$   & 14.32              & $<$0.13                & ... & ...\\
$[$Cl \sc{ii}$]$  & 14.37              & $<$0.14                & ... & ...\\
$[$Ne \sc{iii}$]$ & 15.56              & 2.1 $\pm$ 0.2          &1300$\pm$120 &1200$\pm$120\\
H$_2$ S(1)        & 17.04              & 0.83 $\pm$ 0.05        & 660$\pm$50 & 430$\pm$55\\
$[$Fe {\sc ii}$]$ & 17.94              & $<$0.10                & ... & ...\\
$[$S \sc{iii}$]$  & 18.71              & 0.23 $\pm$ 0.03        & 490$\pm$50 & $<$200\\
$[$Ne \sc{v}$]$   & 24.30              & $<$0.20                & ... & ...\\
$[$O {\sc iv}$]$  & 25.89              & $<$0.35                & ... & ...\\
$[$Fe {\sc ii}$]$ & 25.99              & $<$0.31                & ... & ...\\
\cline{1-5}
$[$O \sc{iii}$]$\tablenotemark{b}  & 0.5007             & 0.45           & ... & 890\\
$[$O \sc{iii}$]$\tablenotemark{c}  & 0.5007             &  \nodata\phn              & ... & 600--1000\\
H$\alpha$\tablenotemark{b}         & 0.6565             & 0.51$\pm$0.03  & ... & 720\\
$[$Fe \sc{ii}$]$\tablenotemark{a}  & 1.257              & 0.228:         & ... & 900:\\
Pa$\beta$\tablenotemark{a}         & 1.282              & 0.115:         & ... & 600:\\
$[$Fe \sc{ii}$]$\tablenotemark{a}  & 1.644              & 0.376:         & ... & 1000:
\enddata
\tablenotetext{a}{\citet{dannerbauer05}}
\tablenotetext{b}{\citet{buchanan06}}
\tablenotetext{c}{\citet{heckman90}}
\tablenotetext{}{{\sc Notes.} --- Line fluxes and 3$\sigma$ upper
limits above the separator line are measured from our Spitzer-IRS
spectrum, those below have been taken from the literature. The 
uncertainties in our line fluxes are the larger of either the
error in the Gaussian fit or half the difference of the two nodding 
positions. The FWHM in column four (five) is from a Gaussian fit to 
the data, before (after) removal of the instrumental profile, 
in quadrature.}
\end{deluxetable}

Of the six detected emission lines, the two neon fine-structure 
lines appear clearly resolved beyond the instrumental resolution 
(500 km s$^{-1}$; $R$=600), most notably the [Ne{\sc ii}] line, at 
an observed FWHM of 1500$\pm$100\,km s$^{-1}$ (Table\,\ref{tab1}). 
The other emission lines have observed FWHM ranging from 490$\pm$50
to 950$\pm$130\,km s$^{-1}$.
In what follows any FWHM values quoted in the text refer to
the true FWHM, i.e. the observed FWHM corrected for the 
instrumental profile.

\begin{figure}
\begin{center}
\includegraphics[angle=0,width=85mm]{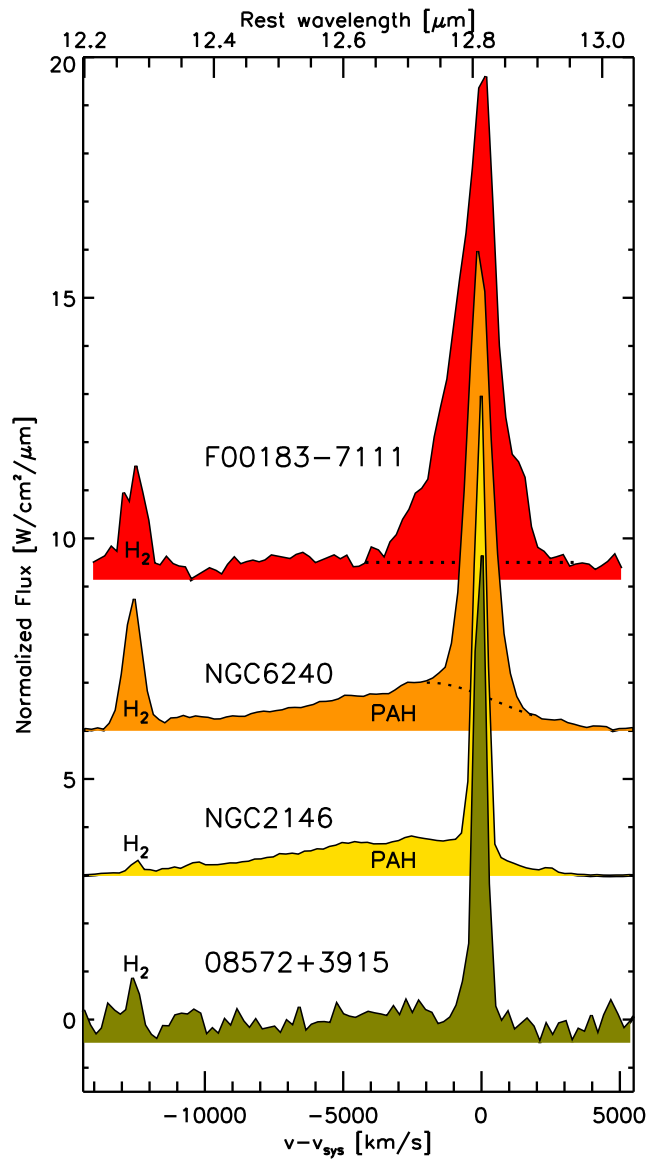}
\caption{Comparison of the continuum-subtracted spectral structure
around the 12.81\,$\mu$m [Ne {\sc ii}] line for four galactic nuclei,
at a spectral resolution $R$=600. The spectral features included are 
the H$_2$ S(2) line at 12.28\,$\mu$m, the 12.7\,$\mu$m PAH emission 
feature (extending from -13,500 to +3,500 km/s) and the 
[Ne {\sc ii}] line. The latter is clearly broadened in the spectra
of IRAS\,F00183--7111 and NGC\,6240. Note that the 12.7\,$\mu$m PAH 
emission feature appears to be absent in the spectra of IRAS\,F00183--7111 
and IRAS\,08572+3915NW
\label{fig2}}
\end{center}
\end{figure}

\begin{figure}
\begin{center}
\includegraphics[angle=0,width=85mm]{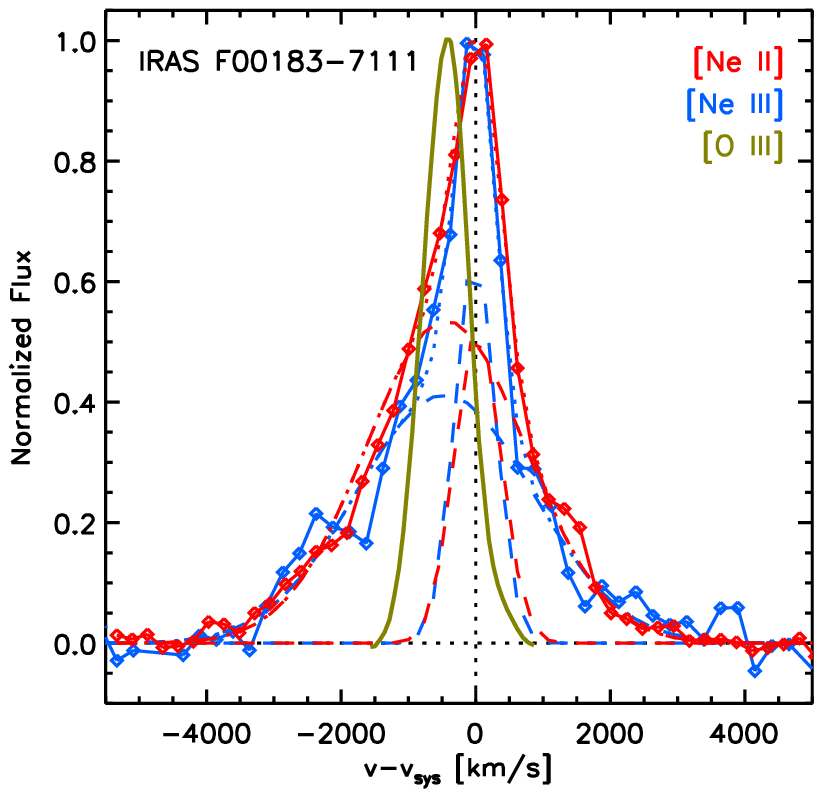}
\caption{Comparison of the galaxy-integrated line profile of 
12.81\,$\mu$m [Ne {\sc ii}] ({\it red}) and 5007\,\AA\ [O {\sc iii}] 
({\it green}) at a position 3$"$ East of the nucleus of 
IRAS\,F00183--7111, where the O$^{2+}$ gas reaches its highest 
blue shift. The spectral resolution of the double-peaked [O {\sc iii}] 
profile ($R$=1900) has been degraded to the resolution of the 
{\it Spitzer}-IRS-SH measurement ($R$=600). Overplotted are the
2-component Gauss fits to the [Ne {\sc ii}] and [Ne {\sc iii}]
line profiles. The individual components are shown as {\it dashed} 
lines, their sums as {\it dotted} lines
\label{fig3}}
\end{center}
\end{figure}

\subsection{Neon line profile analysis}

Extracting a [Ne {\sc ii}] line profile requires careful removal
of the underlying spectral structure. As 
illustrated in Fig.\,\ref{fig2} by the starburst galaxy NGC\,2146
and the ULIRG NGC\,6240, the [Ne{\sc ii}] line is superimposed on 
the 12.7\,$\mu$m PAH emission feature, which arises over a velocity 
range of -13500 to +3500\,km\,s$^{-1}$ relative to the [Ne {\sc ii}] 
line. However, unlike NGC\,2146 and NGC\,6240, the spectrum of 
IRAS\,F00183--7111 shows hardly any PAH emission, with the exception 
of the PAH feature at
11.2\,$\mu$m, which is detected at very low equivalent width. We 
conclude that the remarkable width of the 12.81\,$\mu$m [Ne {\sc ii}] 
line can not be ascribed to inproper subtraction of the underlying 
spectral structure. In fact, the resulting profile is consistent 
with that of the 15.56\,$\mu$m [Ne {\sc iii}] line.

The shape of the [Ne {\sc ii}] line profile is asymmetric, 
with the peak emission arising very close
to systemic velocity. To characterize the asymmetry we have 
decomposed the line profile into two Gaussian components
(Fig.\,\ref{fig1}). The resulting fit is good. We 
find the dominant component to have a FWHM of 2570$\pm$90\,km\,s$^{-1}$,
centered at a blueshift of 400$\pm$50\,km s$^{-1}$ w.r.t. systemic. The 
narrow component has a FWHM of 610$\pm$60\,km\,s$^{-1}$ and a redshift 
of 60$\pm$20\,km\,s$^{-1}$ w.r.t. systemic. 
The fit components are shown overplotted in red onto the 
observed [Ne {\sc ii}] profile in the top-right panel of 
Fig.\,\ref{fig1}. Further fit details are provided in 
Table\,\ref{tab3}. 

Unlike the 12.81\,$\mu$m [Ne {\sc ii}] line, the 15.56\,$\mu$m
[Ne {\sc iii}] line resides in a relatively uncomplicated 
part of the spectrum. We determined its profile by defining 
and subtracting a local third-order polynomial continuum.
We then performed a 2-component decomposition in the same
way as for the [Ne {\sc ii}] line. The narrow component
has a FWHM of 460$\pm$90\,km s$^{-1}$ and is centered at 
the systemic velocity, whereas the broad component has a 
FWHM of 2850$\pm$240\,km s$^{-1}$ and is blue shifted by 
450$\pm$100\,km s$^{-1}$. The 
fraction of the  [Ne {\sc iii}] flux emitted in the broad
component is 3/4, slightly lower than the fraction 
found for the [Ne {\sc ii}] line, 4/5. As a result, the 
[Ne {\sc iii}]/[Ne {\sc ii}] ratios are 0.31$\pm$0.03 
for the broad component and 0.39$\pm$0.08 for the narrow
component.

\subsubsection{Comparison to other line measurements for 
IRAS\,F00183--7111}

The most detailed spectral line study that we are aware
of focusses on the [O {\sc iii}] doublet at 4959 and 5007\,\AA\
and was published by \citet{heckman90}. They observed 
IRAS\,F00183--7111 in a 2$"\times$100$"$ slit along PA 90$\degr$
(see Fig.\,\ref{fig7}) at a spectral resolution $R$=1900.
In Fig.\,\ref{fig3}
we compare the 5007\,\AA\ [O {\sc iii}] profile obtained at 
a position 3$"$ (13\,kpc) East of the nucleus to that of our 
galaxy-integrated spectrum, after degrading
the spectral resolution of the [O {\sc iii}] line from
1900 to 600. According to Fig.\,12 of \citet{heckman90}, at 
this position the centroid of the [O {\sc iii}] line profile
has its highest blueshift (700\,km s$^{-1}$; using our
preferred redshift of 0.328$\pm$0.001) at a FWHM
$\sim$600 km s$^{-1}$, typical for the entire region extending 
from 10\,kpc West to 20\,kpc East of the nucleus. Clearly, the
[Ne {\sc ii}] and [Ne {\sc iii}] emission extends to much 
higher velocities than the [O {\sc iii}] line emission does.

More recently, \citet{buchanan06} published line fluxes and
FWHMs for 11 optical lines observed at $R$=1000 in a 
2$"$$\times$9$"$ slit oriented along PA=329$\degr$
(C.L. Buchanan, private communication). For the H$\alpha$ line, 
which is strongly blended with the $[$ N {\sc ii}$]$ line on 
either side, they infer a FWHM of 720 km s$^{-1}$ 
\citep[compared to 700 km s$^{-1}$ reported by][]{armus89}. 
For the 4959/5007\,\AA\ [O {\sc iii}] doublet, they report 
FWHMs of 800--900 km s$^{-1}$ (Table\,\ref{tab2}), which are
comparable to those given by  \cite{heckman90} in their 
Figure 12, even though the slit orientations in the two 
studies in question differ by 239 degrees (Fig.\,\ref{fig7}).  

We also compared our mid-infrared results to near-infrared
H and K band spectroscopy by \citet{dannerbauer05} obtained
using a 1$"\times$290$"$ slit oriented at PA=90$\degr$ 
(Fig.\,\ref{fig7}). Their data have similar spectral 
resolution as ours. The two detected [Fe {\sc ii}] lines 
at 1.257 and 1.644\,$\mu$m have FWHMs in the range 900--1000 
km s$^{-1}$, while the 1.282\,$\mu$m Pa$\beta$ line appears 
to be unresolved. Again, like in the optical range, none of 
the lines have FWHMs as large as the mid-infrared neon lines 
(see Table\,\ref{tab1}).

\subsection{AGN tracing mid-infrared lines}

The rest wavelength range covered by the {\it Spitzer} 
IRS high-resolution modules includes three high-ionization 
lines commonly used as tracers of AGN activity: 
7.65\,$\mu$m [Ne {\sc vi}] and 14.32\,\&24.32\,$\mu$m 
[Ne {\sc v}] \citep[e.g.][]{sturm02}. None of them are
detected (Table\,\ref{tab1}). The 3$\sigma$ limit on the 
fraction of 8--1000\,$\mu$m infrared power
contained in the 14.32\,$\mu$m [Ne {\sc v}] line is
L$_{\rm [Ne {\sc v}]}$/L$_{\rm IR}$ $<$ 1.3$\times$10$^{-5}$,
which is 1--2 orders of magnitude smaller than the
detections for classical AGNs, but similar to the limits 
derived for other ULIRGs with deep 9.7\,$\mu$m silicate 
features like IRAS\,08572+3915, IRAS\,15250+3609 and 
IRAS\,20100--4156, and for the type-1 ULIRG/LoBAL-quasar 
Mrk\,231 \citep{armus07}. 
For the limit on 7.65\,$\mu$m [Ne {\sc vi}] no ULIRG 
comparison data exists. However, the 3$\sigma$ upper limit 
L$_{\rm [Ne {\sc vi}]}$/L$_{\rm IR}$ $<$ 2.6$\times$10$^{-5}$
is at least an order of magnitude below what
is found for classical AGNs \citep{sturm02}.

\subsection{PAH emission}

The quality of our {\it Spitzer} IRS high-resolution 
spectrum allows us to measure the flux in the 11.2\,$\mu$m
PAH feature more accurately than has been possible from 
the low-resolution spectrum \citep{spoon04}. The latter 
showed strong disagreements between the spectra obtained 
in nod positions 1 and 2 at the short wavelength end of 
IRS Long-Low order 2. The new feature flux, measured from
the IRS SH spectrum is tabulated in Table\,\ref{tab1}
and is 25\% smaller than previously reported by 
\cite{spoon04} based on the low-resolution spectrum.
We also remeasured the upper limit on the 6.2\,$\mu$m 
PAH feature from a re-extracted low-resolution spectrum, 
using the same continuum pivots at 5.65, 6.07 and 
6.50\,$\mu$m to define the local spline continuum
as in \cite{spoon04}. 
Given the clear reduction in noise and artefacts
between IRS pipeline versions 9.1 and 14.2, the new 
3$\sigma$ upper limit on the 6.2\,$\mu$m PAH flux is 
1.9$\times$\,10$^{-21}$ W cm$^{-2}$, which is a factor 
of 2 lower than previously reported by \citet{spoon04}.
Note that we do not attempt to measure
the flux in the 7.7\,$\mu$m PAH band, as this 
emission feature --- if present --- would be strongly 
blended with the silicate feature.

Based on the new PAH fluxes we compute an {\it upper} 
limit on the ratio of 6.2-to-11.2\,$\mu$m PAH of 0.70. 
This is at the {\it low} end of the range found for 
starburst nuclei and ULIRGs in our sample (2$\sigma$ 
below the average ratio of 1.4), but not untypical 
for a sample of IR-faint LINERs 
(L$_{\rm IR}$=10$^{7.83-10.59}$) studied by \cite{sturm06},
and for the spiral arms of M\,51 and the halo of
M\,82 \citep{galliano08}.

The low 6.2-to-11.2\,$\mu$m PAH ratio may indicate
a high ratio of neutral to ionized PAHs, which 
would imply a low ratio of the ionizing radiation 
field to the electron density, and physical conditions 
similar to those in the Galactic reflection nebula 
NGC\,2023 \citep{galliano08}. In IRAS\,F00183--7111
these conditions could plausibly exist outside the
deeply buried nucleus.
Alternatively, the low observed 6.2-to-11.2\,$\mu$m 
PAH ratio for IRAS\,F00183--7111 may indicate
destruction of small PAHs exposed to hard radiation 
from an AGN. This would selectively 
lower the fluxes emitted in the PAH emission bands 
at 3.3, 6.2 and 7.7\,$\mu$m relative to those
at 11.2\,$\mu$m and beyond \citep{smith07}. 
Given, however, the absence of 
detections of high ionization lines, indicative of 
a hard AGN radiation field, this scenario 
may not hold for IRAS\,F00183--7111.

Note that strong extinction on the PAH emitting
region can be ruled out as a third explanation for
the low observed 6.2-to-11.2\,$\mu$m PAH ratio,
as this would have the opposite effect on the
PAH ratio: the attenuation on the 11.2\,$\mu$m PAH 
feature is stronger than the attenuation on the 
6.2\,$\mu$m PAH feature \citep[e.g.][]{chiar06}.

\section{Other ULIRGs showing resolved mid-IR neon lines}

\begin{figure}
\begin{center}
\includegraphics[angle=0,width=85mm]{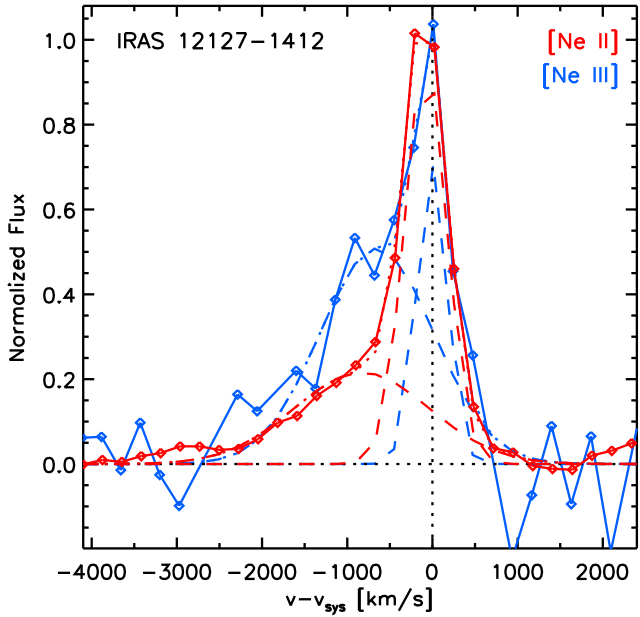}
\caption{Comparison of the galaxy-integrated line profiles 
of 12.81\,$\mu$m [Ne {\sc ii}] ({\it red}) and 15.56\,$\mu$m 
[Ne {\sc iii}] ({\it blue}) for IRAS\,12127-1412NE 
at a spectral resolution of $R$=600. Overplotted are
2-component Gauss fits to the line profiles. The individual
components are shown as {\it dashed} lines, their sums as 
{\it dotted} lines
\label{fig9}}
\end{center}
\end{figure}

The discovery of high velocity [Ne {\sc ii}] emission 
in IRAS\,F00183--7111 raises the question whether there 
are other luminous infrared galaxies that show evidence 
for disturbed (FWHM$\geq$800 km s$^{-1}$) neon gas 
in their {\it Spitzer}-IRS spectra.  If so, do these 
galaxies have properties similar to IRAS\,F00183--7111?
To address this question we have inspected the 
line profiles of the low-ionization emission lines
12.81\,$\mu$m [Ne {\sc ii}] and 15.56\,$\mu$m 
[Ne {\sc iii}] of the 53 ULIRGs in the sample of 
\cite{farrah07}, as well as the archival spectra of 
the ULIRG IRAS\,12127--1412NE and the HyLIRGs 
IRAS\,F09104+4109 and IRAS\,F15307+3252. In our
search we distinguish two classes. One consisting 
of sources with symmetric resolved neon line profiles, 
and another with sources which show asymmetric 
resolved line profiles.

\begin{figure}
\begin{center}
\includegraphics[angle=0,width=85mm]{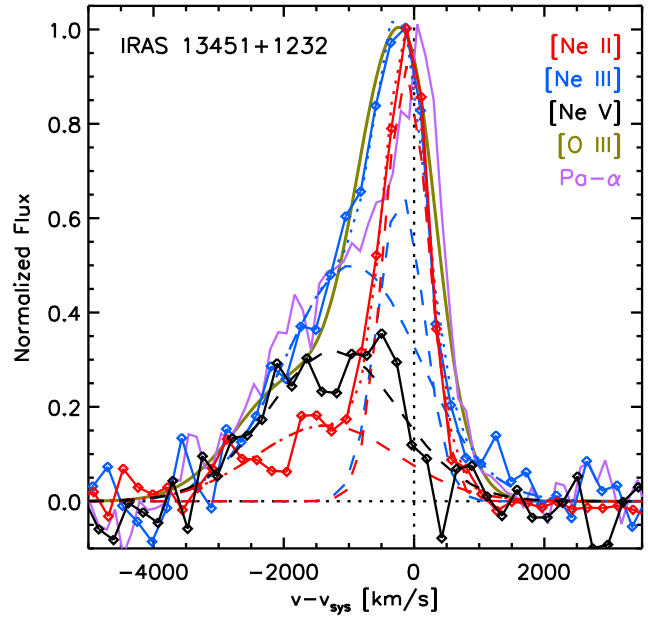}
\caption{Comparison of the galaxy-integrated line profiles 
of 12.81\,$\mu$m [Ne {\sc ii}] ({\it red}), 15.56\,$\mu$m 
[Ne {\sc iii}] ({\it blue}) and 14.32\,$\mu$m [Ne {\sc v}] 
({\it black}) for IRAS\,13451+1232 at a spectral resolution 
of $R$=600. Overplotted are 2-component Gauss fits to the 
neon line profiles. The individual components are shown as 
{\it dashed} lines, their sums as {\it dotted} lines. 
The [Ne {\sc v}] line has been scaled to match the blue wing 
of the [Ne {\sc iii}] line profile. Also shown is the sum 
of the 3-component Gauss fit to the 5007\,\AA\ [O {\sc iii}] 
line ({\it green}) as presented by \cite{holt03}, degraded 
to a resolution of $R$=600, and the Pa$\alpha$ profile 
({\it purple}) obtained by \cite{veilleux97} at $R$=650
\label{fig10}}
\end{center}
\end{figure}

\begin{deluxetable*}{lclcccc}
\tablewidth{0pt} 
\tablecaption{Fit results for mid-infrared neon emission lines\label{tab3}}
\tablehead{\colhead{Feature} & \colhead{Component} & \colhead{$\lambda_{\rm c}$} & 
           \colhead{v$_{\rm c}$} & \colhead{obs. FWHM} & \colhead{true FWHM} & 
           \colhead{Flux}  \\
           \colhead{} & \colhead{} & \colhead{($\mu$m)} & 
           \colhead{(km s$^{-1}$)} & \colhead{(km s$^{-1}$)} & \colhead{(km s$^{-1}$)} & 
           \colhead{(10$^{-21}$\,W\,cm$^{-2}$)}  
          }
\startdata
\cline{1-7}
\multicolumn{7}{c}{IRAS\,00183--7111}\\
\cline{1-7}
$[$Ne \sc{ii}$]$  &n& 12.816  & + 60 $\pm$ 20 &  790 $\pm$ 60 &  610 $\pm$  60&1.4 $\pm$ 0.16 \\
$[$Ne \sc{ii}$]$  &b& 12.796  & -400 $\pm$ 50 & 2620 $\pm$ 90 & 2570 $\pm$  90&4.9 $\pm$ 0.20 \\
$[$Ne \sc{iii}$]$ &n& 15.555  &    0 $\pm$ 30 &  680 $\pm$ 90 &  460 $\pm$  80&0.55$\pm$ 0.09 \\
$[$Ne \sc{iii}$]$ &b& 15.532  & -450 $\pm$ 100& 2890 $\pm$ 240& 2850 $\pm$ 240&1.5 $\pm$ 0.12 \\
$[$Ne \sc{v}$]$   &-& 14.322  &       --      &   --          &  --  &  $<$0.13      \\
\cline{1-7}
\multicolumn{7}{c}{IRAS\,12127--1412NE}\\
\cline{1-7}
$[$Ne \sc{ii}$]$  &n& 12.818  &  -80 $\pm$ 10 &  530 $\pm$ 20 & 175 $\pm$  45& 1.68 $\pm$ 0.08\\
$[$Ne \sc{ii}$]$  &b& 12.799  & -530 $\pm$ 60 & 1740 $\pm$ 100&1670 $\pm$ 100& 1.47 $\pm$ 0.10\\
$[$Ne \sc{iii}$]$ &n& 15.565  &   -7 $\pm$ 40 &  430 $\pm$ 120& $<$200       & 0.16 $\pm$ 0.07\\
$[$Ne \sc{iii}$]$ &b& 15.531  & -650 $\pm$ 200& 1580 $\pm$ 360&1500 $\pm$ 340& 0.42 $\pm$ 0.10\\
$[$Ne \sc{v}$]$   &-& 14.322  &   --          &   --          & --   & $<$0.08     \\
\cline{1-7}
\multicolumn{7}{c}{IRAS\,13451+1232}\\
\cline{1-7}
$[$Ne \sc{ii}$]$  &n& 12.808  & -120 $\pm$ 10 &  790 $\pm$ 40 & 610 $\pm$  50&4.3 $\pm$ 0.30 \\
$[$Ne \sc{ii}$]$  &b& 12.758  &-1290 $\pm$ 215& 2520 $\pm$ 350&2470 $\pm$ 340&2.5 $\pm$ 0.40 \\
$[$Ne \sc{iii}$]$ &n& 15.545  & -200 $\pm$ 25 &  840 $\pm$ 70 & 670 $\pm$  70&1.9 $\pm$ 0.25 \\
$[$Ne \sc{iii}$]$ &b& 15.506  & -950 $\pm$ 85 & 2410 $\pm$ 110&2360 $\pm$ 110&4.1 $\pm$ 0.30 \\
$[$Ne \sc{v}$]$   &b& 14.263  &-1240 $\pm$ 80 & 2360 $\pm$ 190&2300 $\pm$ 190&1.9 $\pm$ 0.15 \\
\enddata
\tablenotetext{}{{\sc Notes.} --- Results from a 2-component gaussian 
decomposition of the 12.81, 14.32 and 15.56\,$\mu$m neon emission line 
profiles into narrow (n) and broad (b) components. 
$\lambda_{\rm c}$ denotes the central wavelength of the component and 
corresponds to a velocity shift v$_{\rm c}$ of the component with 
respect to systemic velocity. The formal error on the velocity 
shift does not include corrections for the uncertainty in the wavelength 
calibration, which is 1/5 of a resolution element (100 km s$^{-1}$).
The FWHM in column five (six) is measured before (after) removal 
of the instrumental profile \citep[500 $\pm$ 60\,km s$^{-1}$;][]{dasyra08}, 
in quadrature}
\end{deluxetable*}

The class of asymmetric resolved [Ne {\sc ii}] and
[Ne {\sc iii}] line profiles consists of only two sources 
besides IRAS\,F00183--7111. 
In all three cases the line asymmetry takes the form 
of a blue wing.

\subsection{IRAS\,12127--1412NE}

One source is IRAS\,12127--1412NE (Fig.\,\ref{fig9}),
the brighter of the two nuclei in this widely separated
LINER system \citep[projected nuclear separation 9$"$,
or 21.9\,kpc;][]{veilleux99}. In the
near-infrared, \cite{imanishi06} classified the 
NE nucleus as a buried AGN based on the absence
of 3.3\,$\mu$m PAH emission and the presence of
3.0\,$\mu$m water ice and 3.4\,$\mu$m hydrocarbon 
absorption bands. The deeply buried nature of the
nuclear power source(s) is confirmed by the presence 
of a deep 10\,$\mu$m silicate feature in the 
{\it Spitzer}-IRS low-resolution spectrum 
\citep{imanishi07}.

On 2008 July 10, we reobserved the source in staring
mode in IRS-SH, using an integration time of 12 cycles 
of 120\,sec. Sky-subtraction was performed using a 
contemporaneous sky background observation of 6 cycles 
of 120\,sec. Data reduction proceeded along the same 
steps as detailed in Sect.\,2.1.

Like for IRAS\,F00183--7111, the asymmetry of the neon
emission lines of IRAS\,12127--1412NE may be characterized 
using a 2-component gaussian fit. For the
[Ne {\sc ii}] line the narrow component is practically
unresolved (FWHM = 175$\pm$45\,km s$^{-1}$) and is
centered close to systemic velocity\footnote{We
infer the systemic velocity of IRAS\,12127--1412NE
from the line centers of the H$_2$ S(1) and S(3) lines
and find 39870$\pm$200\,km s$^{-1}$ (z=0.1330$\pm0.0004$),
in good agreement with the redshift of 0.133 quoted 
by \cite{veilleux02}} (-80$\pm$10\,km s$^{-1}$).
The other component has a FWHM of 1670$\pm$100\,km s$^{-1}$ 
and is centered at a blue shift of 530$\pm$60\,km s$^{-1}$ 
(Fig.\,\ref{fig9}). The fraction of the [Ne {\sc ii}] 
line flux emitted in the broad component is slightly 
smaller than that in the narrow component. This is 
less than for IRAS\,F00183--7111, for which about 4/5 
of the line flux is associated with the broad component. 
We also examined the 15.56\,$\mu$m [Ne {\sc iii}] profile. 
Its shape is not as well constrained as that of the 
[Ne {\sc ii}] line due to a five times lower line flux,
but the profile clearly shows the presence of a blue wing. 
A 2-component Gauss fit results in an unresolved narrow 
component centered at systemic velocity (-7$\pm$40\,km s$^{-1}$)
and a broad component with FWHM of 1500$\pm$340\,km s$^{-1}$
blueshifted by 650$\pm$200\,km s$^{-1}$. We find that the
contribution of the broad component to the [Ne {\sc iii}]
line profile is far stronger than to the [Ne {\sc ii}] 
profile, 3/4 as compared to 1/2 (Fig.\,\ref{fig9}). This
implies a difference in excitation of the narrow 
and broad components. For the narrow component we find
a [Ne {\sc iii}]/[Ne {\sc ii}] ratio of 0.095$\pm$0.04, 
while for the broad component we compute a ratio of 
0.29$\pm$0.08
Unfortunately, a comparison to the line profile of 
the optical [O {\sc iii}] doublet is not possible
as the equivalent widths of these lines in the 
spectrum of \cite{veilleux99} are too low 
for a meaningful analysis of their profiles.

\subsection{IRAS\,13451+1232}

The third ULIRG with asymmetric neon line profiles 
is IRAS\,13451+1232\footnote{We adopt
z=0.12174$\pm$0.00001 as the systemic velocity
for IRAS\,13451+1232. This velocity corresponds
to the  velocity center of the narrow component 
in the 3-component fit to the 4959 \& 5007\,\AA\ 
[O {\sc iii}] doublet in the $R$=1200 spectrum 
of \cite{holt03}. Using this redshift, the line
centers of the H$_2$ S(1) and S(3) lines occur
at 40$\pm30$ and 20$\pm$35\,km s$^{-1}$, respectively.
Note that our adopted redshift is 200\,km s$^{-1}$ 
lower than the value derived from CO observations
\citep{evans99}}, 
also known as PKS\,1345+12 
(Fig.\,\ref{fig10}). The source, which is optically
classified as Seyfert-2 \citep{veilleux97}, comprises 
two nuclei separated by 1.8$"$ (4.3\,kpc; unresolved
at the spatial resolution of {\it Spitzer}-IRS).
Both the low resolution near-infrared 2.6--3.65\,$\mu$m 
\citep{imanishi06} and the {\it Spitzer} mid-infrared
spectrum is virtually featureless, characteristic of 
AGN heated hot dust. At the higher resolving power 
of IRS-SH, emission line profiles of 12.81\,$\mu$m 
[Ne {\sc ii}], 15.56\,$\mu$m [Ne {\sc iii}] and
14.32\,$\mu$m [Ne {\sc v}] can be discerned on
the continuum, the clearest of which is the is 
the [Ne {\sc iii}] line. We fit its profile with 
a broad component of FWHM 2360$\pm$110\,km s$^{-1}$ blue 
shifted by 950$\pm$85\,km s$^{-1}$
and a narrow component of FWHM 670$\pm$70\,km s$^{-1}$ 
blueshifted by 200$\pm$25\,km s$^{-1}$. The broad 
component is responsible for 2/3 of the [Ne {\sc iii}] 
emission, significantly more than for IRAS\,F00183--7111 
but somewhat less than for IRAS\,12127--1412NE.
For the [Ne {\sc ii}] line profile the 2-component
decomposition results in a narrow component with a
FWHM of 610$\pm$50\,km s$^{-1}$ blueshifted by 
120$\pm$10\,km s$^{-1}$ and a broad component with 
FWHM 2470$\pm$340\,km s$^{-1}$ blueshifted by 
1290$\pm$220\,km s$^{-1}$. The broad component 
carries 1/3 of the total [Ne {\sc ii}] line flux,
significantly less than for IRAS\,F00183--7111 and 
IRAS\,12127--1412NE. The third line, the [Ne {\sc v}]
coronal line at 14.32\,$\mu$m, {\it only} has a blue 
shifted component, centered at a -1240$\pm$80\,km s$^{-1}$,
which we fit with a Gaussian profile of FWHM 
2300$\pm$190\,km s$^{-1}$. The presence of [Ne {\sc v}] 
emission only at blueshifted wavelengths is
consistent with the difference in excitation
between the narrow and broad components as measured 
from the [Ne {\sc iii}]/[Ne {\sc ii}] ratio. 
For the narrow component this ratio is 0.44$\pm$0.07, 
while for the broad component it is 1.6$\pm$0.3. 
The [Ne {\sc v}]/[Ne {\sc ii}] ratio for the broad 
component is 0.76$\pm$0.14. 

In Fig.\,\ref{fig10} we 
also show the sum of a 3-component 
fit to the profile of the 5007\,\AA\ [O {\sc iii}]
line, centered on the nucleus \citep{holt03}. The
spectral resolution was degraded from $R$=1200 
to 600 for a proper comparison to our IRS-SH
data. As a result of the degradation, the narrowest 
(FWHM=340$\pm$23\,km s$^{-1}$) and weakest
of the three kinematic components, centered at 
systemic velocity, gets blended with the intermediate 
width component (FWHM=1255$\pm$12\,km s$^{-1}$) 
centered at -402$\pm$9\,km s$^{-1}$.
The broadest, most blueshifted component 
(FWHM=1944$\pm$65\,km s$^{-1}$), centered at 
-1980$\pm$36\,km s$^{-1}$, can however still be
recognized as a separate component. We also
overplot the line profile of Pa$\alpha$ 
\citep{veilleux97} with a narrow component
of FWHM=773\,km s$^{-1}$ and a broad component
of FWHM=2588\,km s$^{-1}$. The overlay suggests that 
the 15.56\,$\mu$m [Ne {\sc iii}] line traces more 
or less the same blue shifted kinematic components 
as the 5007\,\AA\ [O {\sc iii}] and Pa$\alpha$ lines.
The comparison further shows that, unlike
IRAS\,F00183--7111, there is no evidence for
the 5007\,\AA\ [O {\sc iii}] line to cover a 
more limited velocity range than the 12.81\,$\mu$m 
[Ne {\sc ii}] and 15.56\,$\mu$m [Ne {\sc iii}] lines 
(Figs.\,\ref{fig3}\,\&\,\ref{fig10}). The latter
is in agreement with the differences in the 
distribution of the obscuring medium as probed
by the 9.7\,$\mu$m silicate absorption features
(see Section 5.4).

\begin{figure*}
\begin{center}
\includegraphics[angle=0,width=12cm]{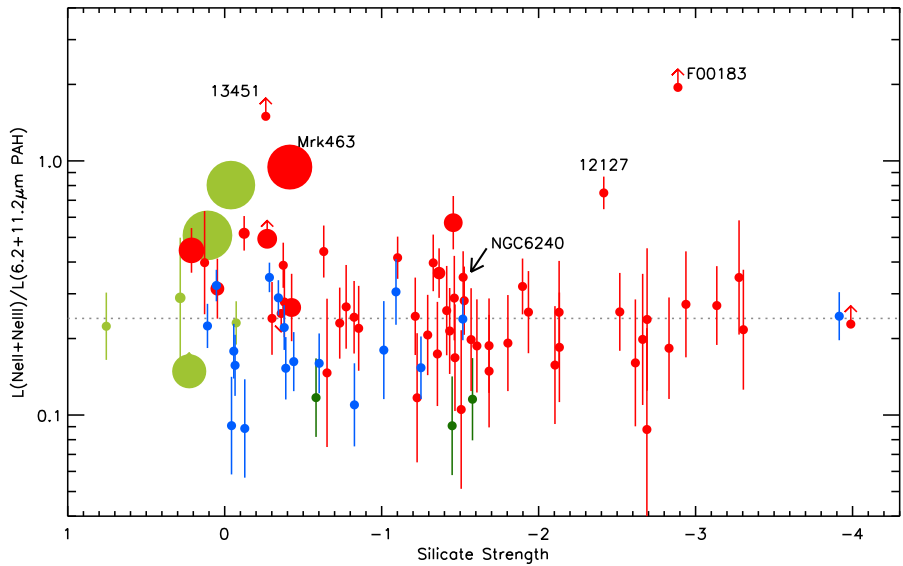}
\caption{
Ratio of the combined [Ne {\sc ii}]+[Ne {\sc iii}] luminosity and
the combined PAH(6.2\,$\mu$m)+PAH(11.2\,$\mu$m) luminosity as a 
function of silicate strength \citep{spoon07}. The symbol size scales 
linearly with the [Ne {\sc v}]/[Ne {\sc ii}] ratio above a threshold 
ratio of 0.4. The largest symbol corresponds to a 
[Ne {\sc v}]/[Ne {\sc ii}] ratio of 2.2.
The horizontal {\it dotted} line indicates the average
neon/PAH ratio for {\it low-excitation} ULIRGs. IRAS\,F00183--7111
is more than a factor 8 above this value, indicating that it is 
overluminous in its neon lines compared to its PAH emission.
The same is true to a lesser extent for the ULIRGs IRAS\,12127--1412NE
and IRAS\,13451+1232. Galaxy types are labeled as follows:
ULIRGs: {\it red}, starburst galaxies: {\it blue}, Seyfert-2 galaxies:
{\it dark green}, Seyfert-1 galaxies and QSOs: {\it light green}
\label{fig5}}
\end{center}
\end{figure*}

\section{Discussion}

\subsection{AGN activity in IRAS\,F00183--7111}

Observations at various wavelengths suggest that 
IRAS\,F00183--7111 derives a substantial fraction of its
bolometric power from accretion onto a central black hole. 
Especially important in this respect is the detection 
by \citet{nandra07} of a 6.7\,keV Fe\,K$\alpha$ line 
(Fe {\sc xxv}) with a large equivalent width, indicating 
the presence of a Compton-thick AGN in IRAS\,F00183--7111.
Another clear indication is a higher than expected radio 
power based on the radio-far-infrared correlation for starburst 
galaxies \citep{roy97}, as well as unpublished radio 
interferometric data showing the nucleus to be unresolved 
on the 200\,pc scale (R.P. Norris, 2004, private communication).

The non-detection of several common mid-infrared AGN tracers
in our {\it Spitzer}-IRS high resolution spectrum may 
hence indicate that the AGN coronal line region (CLR) is 
small or absent or, alternatively, that our line of sight 
to the AGN CLR is strongly obscured. The latter would be 
in agreement with the detection of a deep 9.7\,$\mu$m 
silicate absorption feature \citep[Fig\,\ref{fig4}]{spoon04}, 
which is unusual for classical seyfert galaxies and
quasars \citep{spoon07,hao07}. Even stronger obscuration,
along all lines of sight, is suggested by \citet{lutz04}, 
who concluded from a study of the 4.7\,$\mu$m CO gas 
absorption band that the strong detection of this feature 
in IRAS\,F00183--7111 \citep{spoon04} and its non-detection 
in local AGNs observed with ISO indicates that the central 
power source in IRAS\,F00183--7111 must be fully covered 
rather than blocked by an edge-on torus.

\subsection{Excitation conditions in the outflow of IRAS\,F00183--7111}

To gain insight into the processes responsible for
the presence of high velocity gas in IRAS\,F00183--7111
we compare IRAS\,F00183--7111 to the only other 
two ULIRGs in our sample that show pronounced blue wings 
in their 12.81\,$\mu$m [Ne {\sc ii}] and 15.56\,$\mu$m 
[Ne {\sc iii}] line profiles: IRAS\,12127-1412NE 
(Fig.\,\ref{fig9}) and IRAS\,13451+1232 (Fig.\,\ref{fig10}). 

Clear evidence for unusual excitation conditions in 
IRAS\,F00183--7111 emerge if we compare the PAH emission 
and low-ionization neon line emission for  AGNs, ULIRGs 
and starburst galaxies from the various {\it Spitzer}-IRS 
GTO programs, (U)LIRGs from {\it Spitzer} program 1096
and IRAS\,12127--1412NE. The PAH luminosity 
should track the low-ionization neon line luminosity to 
within a factor two, if predominantly excited in 
star forming regions. This was indeed 
found to be the case for the sample of ULIRGs published 
by \cite{farrah07}. In Fig.\ref{fig5} we plot the ratio 
of these two quantities (the combined 
[Ne {\sc ii}]+[Ne {\sc iii}] 
luminosity and the combined luminosity of the 6.2 and 
11.2\,$\mu$m PAH features) as a function of increasing 
apparent depth of the 9.7\,$\mu$m silicate feature 
\citep[i.e. the negative of the silicate strength, S$_{sil}$;][]{spoon07}.
The symbol size scales linearly with the 
[Ne {\sc v}]/[Ne {\sc ii}] ratio above a threshold
value of 0.4. This distinguishes ULIRGs with a strong
AGN contribution to the [Ne {\sc ii}]+[Ne {\sc iii}] 
luminosity \citep[e.g. Mrk\,463;][]{farrah07} from 
those without. As expected, for the sources lacking 
a strong AGN signature the average value of neon/PAH 
has a reasonably low scatter of a factor 
1.4\footnote{In this computation we excluded 
the outliers IRAS\,F00183--7111, IRAS\,12127--1412NE 
and IRAS\,13451+1232. Keeping them included would increase
the scatter to a factor 1.7}, in agreement with the 
scaling relation between neon and PAH flux for ULIRGs 
found by \cite{farrah07}. 
The notable exception is IRAS\,F00183--7111, which 
appears to be at least 8 times overluminous in the 
neon lines compared to its combined PAH luminosity. 
Another exception in the S$_{sil}$$<$--1.5 regime is 
IRAS\,12127--1412NE, which is overluminous in neon/PAH 
by a factor of 3. At S$_{sil}$$>$--1.5, the situation
is less straightforward. In this regime there are 
several sources with large [Ne {\sc v}]/[Ne {\sc ii}]
ratios for which the AGN contributes significantly
to the [Ne {\sc ii}]+[Ne {\sc iii}] luminosity, thereby
raising the neon/PAH ratio (Fig.\ref{fig5}). 
The only ULIRG in this regime which has an elevated 
neon/PAH ratio despite a low [Ne {\sc v}]/[Ne {\sc ii}] 
ratio of 0.28 is IRAS\,13451+1232,
at $>$6 times the average neon/PAH ratio (Fig.\ref{fig5}).
This brings the number of ULIRGs with unusual excitation 
conditions to three. 
Since these same three ULIRGs are also the only ULIRGs 
in our sample that show pronounced blue wings in their 
12.81\,$\mu$m [Ne {\sc ii}] and 15.56\,$\mu$m [Ne {\sc iii}] 
line profiles, it seems plausible that the high
neon/PAH ratios for these sources could be the result of
additional neon line emission contributed by these blue 
wings rather than unusually weak PAH emission in these 
three sources.

\begin{figure}
\begin{center}
\includegraphics[angle=0,width=85mm]{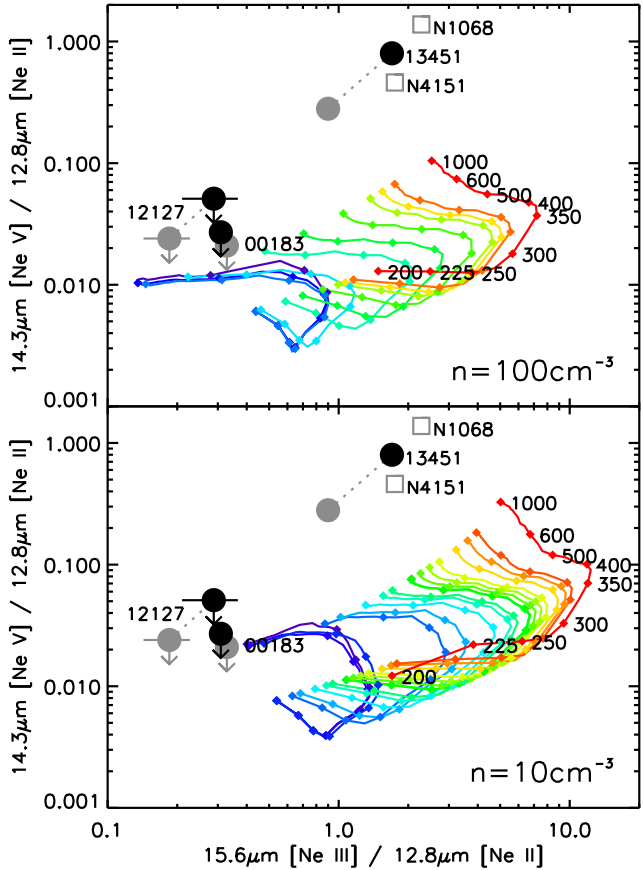}
\caption{Mid-infrared diagnostic diagram for the line ratio
15.56\,$\mu$m [Ne {\sc iii}]\,/\,12.81\,$\mu$m [Ne {\sc ii}] versus
14.32\,$\mu$m [Ne {\sc v}]\,/\,12.81\,$\mu$m [Ne {\sc ii}]. Plotted
are predicted line ratios as computed by MAPPINGS version IIIq
\citep{allen08} for combined precursor+shock models at solar 
metallicity at a gas density n=100\,cm$^{-3}$ (upper panel) and 
n=10\,cm$^{-3}$ (lower panel). Different colors (ranging from
{\it purple} to {\it red}) denote different magnetic field 
strengths, ranging from 10$^{-3}$ to 100\,$\mu$G. 
Shock velocities range from 200 to 1000\,km s$^{-1}$ and are 
labeled along the {\it red} curve.
Overlaid are the positions of IRAS\,F00183--7111, 
IRAS\,12127-1412NE and IRAS\,13451+1232 for the integrated
fluxes ({\it gray disks}) and for just the blue shifted component 
({\it black disks}). {\it Gray squares} mark the positions
of the nuclear spectra of the classical AGNs NGC\,1068 and 
NGC\,4151 \citep{sturm02}
\label{fig11}}
\end{center}
\end{figure}

In Fig.\,\ref{fig11} we show a mid-infrared diagnostic 
diagram based on the 12.81\,$\mu$m [Ne {\sc ii}],
15.56\,$\mu$m [Ne {\sc iii}] and 14.32\,$\mu$m [Ne {\sc v}]
lines. In the diagram we plot shock excitation models
from the MAPPINGS IIIq library \citep{allen08} 
for solar metallicity gas with densities n=100\,cm$^{-3}$
(upper panel) and n=10\,cm$^{-3}$ (lower panel). 
The model tracks plotted include the ionizing effects 
from UV and soft X-ray photons in the shock cooling zone 
on the gas ahead of the shock front. Overplotted are
the positions of IRAS\,F00183--7111, IRAS\,12127+1412NE
and IRAS\,13451+1232 from our Spitzer observations. 
Black filled circles indicate the results for just 
the broad blue-shifted emission line components, 
the gray versions denote the results based on the 
velocity-integrated line fluxes (i.e. the narrow and 
broad components combined).
For IRAS\,F00183--7111 and IRAS\,12127+1412NE the low
observed [Ne {\sc iii}]/[Ne {\sc ii}] line ratios are 
consistent with high-velocity shocks (v$>$500 km s$^{-1}$) 
in high-density environments (n$>$100\,cm$^{-3}$) with 
low magnetic field strengths ($<$1\,$\mu$G). At lower
gas densities (n$=$10\,cm$^{-3}$), only the highest
velocity shocks (v=1000 km s$^{-1}$) at the lowest
magnetic field strengths ($<$0.1\,$\mu$G) are consistent 
with the observed [Ne {\sc iii}]/[Ne {\sc ii}] line 
ratios. Below this density the observed line ratios
cannot have an origin in shock emission. Given the 
absence of density measurements we cannot rule out 
either scenario.
However, as we will argue later, if the neon line
emission originates at the base of the optical 
outflow, gas densities of n$>$100\,cm$^{-3}$ may not
be unreasonable.
In IRAS\,13451+1232 it is the high observed 
[Ne {\sc v}]/[Ne {\sc ii}] ratio which appears to
rule out shocks of any velocity and density
as the source of excitation for the neon lines. 
For the latter source, exposure of the blue shifted
gas to direct radiation from the AGN seems a likely 
explanation for the presence of [Ne {\sc v}] emission
at the observed ratio to [Ne {\sc ii}]. This explanation
is supported by the proximity of the blue shifted component 
of IRAS\,13451+1232 to the positions of the classical 
AGNs NGC\,1068 and NGC\,4151 in Fig.\,\ref{fig11}.

\subsection{Strong obscuration on the 
outflow in IRAS\,F00183--7111}

In Sect.\,3.3 we analyzed the velocity structure of
the 12.81\,$\mu$m [Ne {\sc ii}] and 15.56\,$\mu$m
[Ne {\sc iii}] lines in IRAS\,F00183--7111 and 
compared the results to those of studies of its
optical and near-infrared forbidden lines. 
This revealed that the mid-infrared neon emission 
lines have strongly blue shifted components
not detected in both the optical 
5007\,\AA\ [O {\sc iii}] and near-infrared 
1.644\,$\mu$m [Fe {\sc ii}] line profiles.
Given the mid-infrared spectral evidence 
for strong obscuration in the nucleus of 
IRAS\,F00183--7111 \citep{spoon04}, the 
simplest explanation for the absence of strongly 
blue shifted gas components in the optical
and near-infrared emission lines is extinction.
We estimate the required extinction by assuming 
line emission at any wavelength $\lambda$
to be effectively suppressed by a specific 
extinction A($\lambda$)$\geq$2. The level 
of screen extinction required to obscure the
high-velocity components of the 5007\,\AA\ 
[O {\sc iii}] line then amounts to 
A$_{\rm 5007 \AA\ }$ = A$_V$$>$2.
If the same screen is responsible for obscuring
the high-velocity components of the 1.644\,$\mu$m 
[Fe {\sc ii}] line, the required extinction
increases to A$_V$$>$10 (A$_{1.64 \mu m}$$>$2). 
Then to still be able to see the strongly blue
shifted 12.81\,$\mu$m [Ne {\sc ii}] emission through
this screen, the maximum extinction is constrained 
to A$_V$$<$50 (A$_{12.81 \mu m}$$<$1). This implies 
a column density of the obscuring screen of 
N$_{\rm H}$ = 2--10 $\times$ 10$^{22}$ cm$^{-2}$, 
assuming N$_{\rm H}$ = 1.9 $\times$ 10$^{21}$ 
$\times$ A$_V$. 

In the absence of spatial information for the
mid-infrared neon line emission, the location of 
the screen blocking the optical and near-infrared
line emission cannot be determined with any certainty.
However, it is plausible for the 
screen to be situated at the base of the outflow 
traced in 5007\,\AA\ [O {\sc iii}]. The observed 
strong decrease in blueshift of the 5007\,\AA\ 
[O {\sc iii}] line from 3$"$ to 0$"$ East of the 
nucleus \citep[Fig.\,12 of ][]{heckman90} may 
hence indicate that the material that obscures
the most highly disturbed optical (but not 
mid-infrared) line emission is located within 
the inner 10\,kpc.

\subsection{The evolutionary state of IRAS\,F00183--7111}

IRAS\,F00183--7111 is a clear outlier
in a mid-infrared diagnostic diagram which combines 
the equivalent width of the 6.2\,$\mu$m PAH emission 
feature and the silicate strength in the so-called
`fork' diagram \citep{spoon07}. In this diagram
most ULIRGs are found along one of two branches. 
One branch connects starburst-dominated mid-infrared 
spectra (e.g. NGC\,7714 and M\,82) to spectra of
deeply buried central sources, characterized by very 
deep silicate features and the (virtual) absence 
of PAH emission (e.g. IRAS\,08572+3915). The other
branch connects starburst-dominated sources with
featureless hot dust-dominated sources, which are
mostly AGN dominated.

IRAS\,F00183--7111 is located in the scarcely 
populated region in between the tips of the two 
branches and has IRAS\,12127+1412NE as its nearest 
neighbor. Their location, away from other deeply 
obscured sources, either implies a more than 10
times lower 6.2\,$\mu$m PAH equivalent width than 
other ULIRGs with similar silicate strength, or a 
less pronounced silicate feature than other deeply
obscured ULIRGs. \cite{spoon07} have proposed that 
sources in between the two branches may be AGN
evolving from a fully covered obscuring geometry 
to an incomplete or clumpy obscuring geometry,
with a decrease in apparent depth of the silicate 
feature as a result. In this scenario, the detection 
of strongly blue shifted neon line emission would 
then trace a unique phase in the (partial) 
disruption of the obscuring medium around the 
powerful AGN. Alternatively, the isolated positions
of IRAS\,F00183--7111 and IRAS\,12127+1412NE in 
the fork diagram may be the result of unusual lines 
of sight into these systems. The unusual mid-infrared 
excitation conditions and kinematics may then be 
associated to that.

If sources with mid-infrared kinematic properties 
and neon/PAH ratios similar to IRAS\,F00183--7111 
and IRAS\,12127+1412NE are indeed AGN evolving from 
full to incomplete or clumpy obscuration, there
should be observational characteristics linking 
them. However, the only ULIRG in our sample with
these same mid-infrared properties, IRAS\,13451+1232, 
has a mid-infrared continuum spectrum which is 
very different from the 
other two sources, displaying only a weak silicate 
absorption feature on an otherwise featureless 
power law spectrum. This places it firmly on 
the AGN branch in the fork diagram, far off
from the other two sources. Evidence supporting 
an unusual evolutionary stage is presented by 
\cite{holt03}. Based on the shape 
of its radio SED and a kinematic study of its 
optical emission line spectrum, they conclude
that IRAS\,13451+1232 must be a young compact 
radio source whose radio jets are expanding 
through a dense nuclear dust shell, entraining 
gas and dust from this shell. They infer the 
reddening to be highest toward the broadest 
(most disturbed) most blue shifted kinematic 
component in their optical line study, reminiscent 
of our findings for IRAS\,F00183--7111 using
optical, near and mid-infrared line profiles.

\section{Conclusions}

We have discovered high-velocity ionized gas with 
velocities ranging from -3500 to +3000 km s$^{-1}$ 
as part of a nuclear outflow from 
the ULIRG IRAS\,F00183--7111. 
The gas is traced by the mid-infrared 12.81\,$\mu$m 
[Ne {\sc ii}] and 15.51\,$\mu$m [Ne {\sc iii}] lines,
which are dominated by blue shifted components.
Optical and near-infrared spectroscopic studies show
no evidence for similar high-velocity gas components
in forbidden lines at shorter wavelengths.
We interpret this as the result of strong extinction
(A$_V$=10--50; A$_{12.81 \mu m}$$<$1) on the blue shifted 
gas, identifying the base of the optical outflow 
($<$3$"$ East) as its most likely origin.

Unusual excitation conditions in IRAS\,F00183--7111 
are further implied by a comparison of the combined 
[Ne {\sc ii}]  and [Ne {\sc iii}] luminosity and 
the combined 6.2 and 11.2\,$\mu$m PAH luminosity 
for a large sample of AGNs, ULIRGs and starburst 
galaxies. This reveals the neon line emission in 
IRAS\,F00183--7111 to be at least 8 times stronger 
compared to its PAH emission than the typical ratio
for the sample.

Two other ULIRGs in our sample of 56 display 
mid-infrared excitation and kinematic properties 
similar to IRAS\,F00183--7111: IRAS\,12127--1412NE 
and IRAS\,13451+1232. Both sources have elevated 
neon/PAH ratios and exhibit pronounced blue wings 
in their 15.56\,$\mu$m [Ne {\sc iii}] line profiles. 
Wings are also present in their 12.81\,$\mu$m
[Ne {\sc ii}] profiles, but are less pronounced,
indicating blue shifted gas to have a higher 
[Ne {\sc iii}]/[Ne {\sc ii}] ratio than gas closer
to systemic velocity. IRAS\,13451+1232 even shows
strongly blue shifted 14.32\,$\mu$m [Ne {\sc v}] 
emission. 

While for IRAS\,13451+1232 the observed 
[Ne {\sc iii}]/[Ne {\sc ii}] and 
[Ne {\sc v}]/[Ne {\sc ii}] ratios indicate exposure
of the blue shifted gas to direct radiation from
the AGN, for IRAS\,F00183--7111 and IRAS\,12127--1412NE
the observed ratios are consistent with an origin 
in fast shocks (v$>$500 km s$^{-1}$) in an 
environment with gas densities n$>$10\,cm$^{-3}$.

IRAS\,F00183--7111 and IRAS\,12127--1412NE are 
clear outliers in 
the mid-infrared diagnostic diagram of silicate
strength and 6.2\,$\mu$m PAH equivalent width, the
so-called `fork' diagram \citep{spoon07}. The sources 
are found in between the two branches which separate
deeply buried nuclei and sources with incomplete
or clumpy obscuring geometries. \cite{spoon07}
have proposed that ULIRGs in between the tips of
these branches may be AGN in the process of 
disrupting their fully covered geometry before 
settling on the AGN branch. IRAS\,13451+1232
would then already be there, given its nearly
featureless mid-infrared power law spectrum.

The results from our mid-infrared kinematic and 
excitation study are consistent with the above 
evolutionary scenario in that the strongly blue
shifted [Ne {\sc ii}] and [Ne {\sc iii}] emission
(from shocks) may trace the (partial) disruption 
of the obscuring medium around buried AGNs in
IRAS\,F00183--7111 and IRAS\,12127--1412NE.
The detection of strongly blue shifted 
[Ne {\sc v}] emission in IRAS\,13451+1232 may 
then indicate the lifting of the obscuration around
the AGN coronal line region to be further advanced
in this ULIRG  than in IRAS\,F00183--7111 and 
IRAS\,12127--1412NE, where this line is undetected.

\appendix

\subsection{Molecular hydrogen lines}

\begin{figure}
\begin{center}
\includegraphics[angle=0,width=85mm]{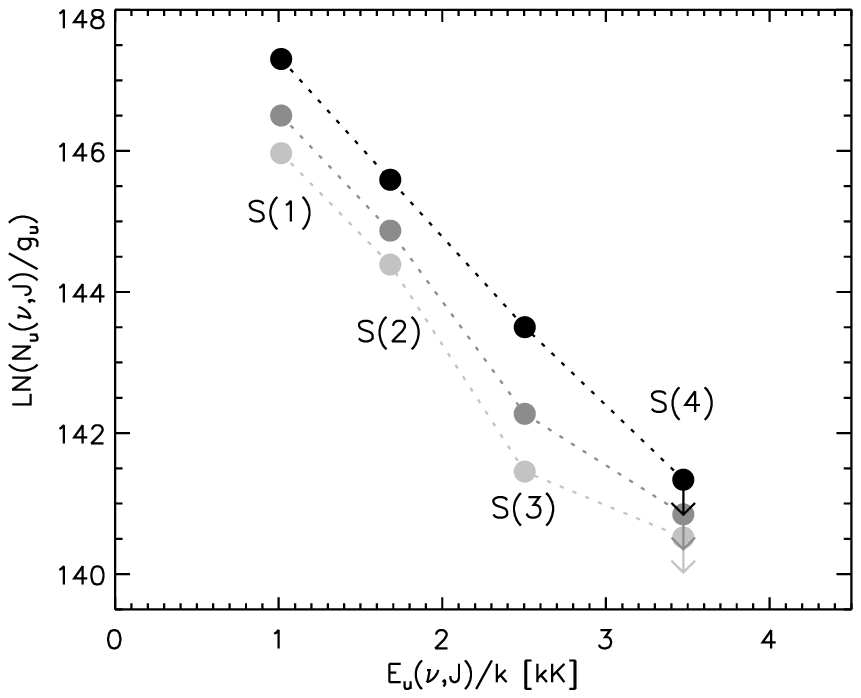}
\caption{Excitation diagram for warm molecular hydrogen in IRAS\,F00183--7111 
based on the line fluxes for the pure-rotational transitions S(1)--S(4).
Different shades of gray are used to distinguish different levels of
dereddening: {\it black}: A$_{\rm V}$=25, {\it dark gray}: A$_{\rm V}$=10, 
{\it light gray}: A$_{\rm V}$=0. Error bars are too small to be
visible
\label{fig6}}
\end{center}
\end{figure}

Our measurements of the S(1), S(2), S(3) and S(4) pure 
rotational molecular hydrogen lines allow us to investigate 
the properties of warm molecular hydrogen in IRAS\,F00183--7111.
Unfortunately, a measurement of the important S(0) line 
is lacking as it is outside the IRS-LH bandpass.
In Fig.\,\ref{fig6} we show the excitation diagram
for molecular hydrogen for the four measured lines.
Given the placement of the S(3) line at the peak of 
the silicate absorption feature, we can attempt an 
extinction correction on the four data points under
the assumption that a) the excitation temperature 
should be constant or increase monotonically 
with increasing upper level energy E$_u$($\nu$,J),
b) that the ortho-to-para ratio is 3 (i.e. assuming 
LTE), and c) that the cooler gas probed by the S(1) 
and S(2) lines is more or less co-spatial with the
warmer gas probed by the S(3) and S(4) lines. 
Adopting the local ISM extinction law 
of \citet{chiar06}, we then find that the extinction 
on the H$_2$ lines ranges between A$_V$=15 and 25.
This is far lower than the extinction on the
continuum source, which is A$_V$$>$54 based on the
apparent depth of the 9.7\,$\mu$m silicate feature 
\citep[$\tau_{9.7}$=3.0;][]{spoon06} and
A$_V$/$\tau_{9.7}$=18 \citep{roche84}.
%
%
Based on the two detected ortho-H$_2$ lines,
it is possible to infer an excitation temperature
for the warm molecular hydrogen. The value ranges
from 330 to 390\,K for extinction corrections ranging 
from A$_V$=0 to 25. 
Assuming all H$_2$ levels to be in LTE at 330--390\,K, 
we derive a total warm molecular hydrogen mass of 
8--20\,$\times$\,10$^8$ M$_{\odot}$. This
value becomes a lower limit if the excitation 
temperature based on the upper level populations
of the S(0) and S(1) lines is lower than those
based on the S(1) and S(3) lines -- which is
usually the case \citep[e.g.][]{higdon06}.
Unfortunately, we cannot determine the warm to total 
molecular hydrogen fraction and compare this to other 
ULIRGs, as no millimeter CO line measurement exists 
for IRAS\,F00183--7111 from which to infer the cold
molecular hydrogen mass.

\subsection{Non-detection of absorption bands of C$_2$H$_2$ 
and HCN gas}

Several ULIRGs show absorption bands of C$_2$H$_2$ and HCN gas
in their 13.7--14.1\,$\mu$m {\it Spitzer} IRS-SH spectra
\citep{spoon05,lahuis07}. Detections are especially strong 
among ULIRGs with pronounced silicate absorption features,
like IRAS\,20100--4156 and IRAS\,15250+3609.
Far weaker detections are reported for ULIRGs with AGN-like 
spectra, like IRAS\,05189--2524, IRAS\,01003--2238 and 
Mrk\,231 \citep{lahuis07}. Detections at these levels 
require spectra to have signal-to-noise ratios of at least
100 in their continua. 
The IRS-SH spectrum of IRAS\,F00183--7111 does not show
any indication for the presence of either feature, despite 
a continuum S/N of 70. This translates into upper
limits on the C$_2$H$_2$ and HCN gas column densities of
N(C$_2$H$_2$)$<$1.2$\times$10$^{16}$\,cm$^{-2}$ and 
N(HCN)$<$3.0$\times$10$^{16}$\,cm$^{-2}$ for an adopted
excitation temperature of 300\,K (F. Lahuis, private 
communication).


\acknowledgements

The authors thank Jan Cami, Helmut Dannerbauer, Brent Groves, Lei Hao, 
Timothy Heckman, Lisa Kewley, Dieter Lutz and Xander Tielens for 
discussions, Catherine Buchanan, Joanna Holt and Sylvain Veilleux for 
sharing their published spectra, and Mark G. Allen for providing
MAPPINGS shock models ahead of publication. We also would like to 
thank the slow but dilligent referee for comments which helped to 
strengthen the paper.
Support for this work was provided by NASA. This research has made
extensive use of the NASA/IPAC Extragalactic Database (NED) which is
operated by the Jet Propulsion Laboratory, California Institute of
Technology, under contract with NASA.


\end{document}